\documentclass{article}

\usepackage{arxiv}

\usepackage[utf8]{inputenc} 
\usepackage[T1]{fontenc}    
\usepackage{hyperref}       
\usepackage{url}            
\usepackage{booktabs}       
\usepackage{amsfonts}       
\usepackage{nicefrac}       
\usepackage{microtype}      
\usepackage{lipsum}

\ifpdf \usepackage[pdftex]{graphicx} \pdfcompresslevel=9
\else \usepackage[dvips]{graphicx} \fi

\usepackage{t1enc,dfadobe}

\usepackage{egweblnk}
\usepackage{cite}

\usepackage{nicefrac}
\usepackage{amsmath}
\usepackage{bm}
\usepackage{amsfonts}
\usepackage{dsfont}
\usepackage[textsize=tiny]{todonotes}
\usepackage{subfigure}
 \usepackage{comment} 
 
 \usepackage[usenames,dvipsnames]{pstricks}
 \usepackage{epsfig}
 \usepackage{pst-grad} 
 \usepackage{pst-plot} 
 \usepackage[space]{grffile} 
 \usepackage{etoolbox} 
 
\usepackage{xcolor,colortbl}
\definecolor{turquoise}{rgb}{0.19, 0.84, 0.78}

\newcolumntype{M}[1]{>{\centering\arraybackslash}m{#1}}
 
 \makeatletter 
 \patchcmd\Gread@eps{\@inputcheck#1 }{\@inputcheck"#1"\relax}{}{}
 \makeatother

\DeclareMathAlphabet\mathbfcal{OMS}{cmsy}{b}{n}

\title{An Introduction to the Deviatoric Decomposition in Three-Dimensions Based on
  a Recursive Formula}

\author{
  Anja Barz \\
  Department of Computer Science\\
  Leipzig University \\
  Germany
  \texttt{} \\
   \And
  Chiara Hergl \\
  Institute for Software Technology, \\
  German Aerospace Center (DLR), Cologne \\
  Germany
  \texttt{} \\
   \And
  Gerik Scheuermann \\
  Department of Computer Science\\
  Leipzig University \\
  Germany
}

\usepackage{color}

\begin{document}
\maketitle

\begin{abstract}
Higher-order tensors appear in various areas of mechanics as well as
  physics, medicine or earth sciences. 
As these tensors are highly complex, most are not well understood.
Thus, the analysis and the visualization process form a highly
  challenging task. 
In order to solve this problem a suitable tensor decomposition for
  arbitrary tensors is desirable. 
Since many real-world examples are described in the three-dimensional
  space, this paper will also focus on this space. 
Tensors of order zero and one are well understood and for
  second order tensors the well-known spectral decomposition exists.
However, there is no standard decomposition for tensors of order higher than two.
This work will focus on summarizing facts about the deviatoric
  decomposition and on its calculation in a recursive way.
It decomposes a tensor of arbitrary order into a sum
  of orthogonal, irreducible parts. 
Thus, into parts that are orthogonal and invariant under rotation.
These irreducible parts are based on so-called deviators.
The calculations and results of the deviatoric decomposition are not
  broadly understood in the engineering community. 
To change this situation, information about this decomposition from a
  range of literature sources was gathered and will be summarized and
  explained in this work to make further analysis possible. 
Different applications and examples for often used physical tensors will
  also be given. 
The goal of this work is to give other researchers the chance to also 
  use this knowledge. 

\end{abstract}


\section{Introduction}
Tensors are mathematical objects, which appear in numerous applications in physics,
  medicine or earth sciences.
However, general $n$th-order tensors having $3^n$ components,
  makes their evaluation a rather difficult task,
  since this number becomes very large quite quickly.
It is preferable to divide a tensor into smaller parts, 
  whose analysis can be understood and worked more effectively.
For certain second order tensors, which can be represented by matrices,
  the spectral decomposition is a useful tool to do so.
Nevertheless, for tensors of order higher than two, which are not equipped with
  particular symmetry types, not many tools are broadly known.

The main focus of this work is to gather information
  about an orthogonal, irreducible 
  decomposition of a general $n$th-order tensor
  into a specific number of independent totally 
  symmetric and traceless tensors, called deviators.
The advantages of this deviatoric decomposition are powerful.
Not only is it unique, but the irreducible parts are 
  orthogonal and invariant under rotation.
Further can this decomposition be calculated recursively. 
Another important benefit concerns the deviatoric tensors.
They can be represented by an, except for an even number of sign changes, unique set of vectors and one scalar. 
This decomposition was summarized by, for example, Hergl et al.
  \cite{hergl2020introduction} and further analyzed by Hergl \cite{herglanalysis}.

The fundamental ideas behind the mathematical concept of the
  deviatoric decomposition are based on group theory
  and were summarized by Backus \cite{backus1970geometrical}.
A deeper understanding of the connection between group theory and tensors
  was also examined by Hamermesh \cite{hamermesh2012group}.
Zou et al. \cite{zou2001orthogonal} then introduced a recursive formula,
  which is based on tensors of lower order,
  to calculate this decomposition
  without needing the group theoretical background.
Formulating this recursive calculation was made possible due to some preliminary
  considerations done by Zheng et al.\cite{zheng2000irreducible}.
This work focuses on summarizing the most important facts about this
  powerful decomposition and provides a closer look at the proof of 
  numerous statements. 
The following structure is used for this work.
At first a general overview of the used tensor algebra is given. 
Important features used in this work are defined and notations are explained.
In Chapter 3 some preliminary considerations, 
  concerning tensors belonging to the space 
  $\mathcal{V} \otimes \mathcal{D}^{(n)}$ are made.
These $(n+1)$th-order tensors are also totally symmetric
  and traceless regarding the indices $(i_1 \cdots i_n)$.
It is important for the following work to identify a 
  suitable decomposition of tensors in this space.
The key to find the decomposition is to
  analyze linear combinations of certain permutations.
Using the tracelessness of $(n)$ indices, the final
  form is created.
In Chapter 4 the main results of this work are confirmed.
At first the existence of the deviatoric decomposition is
  proven by mathematical induction.
During the inductive step the important feature of being recursive becomes 
  visible.
Next, an exact computation of the number $J_s^n$ is given, which
  describes the number of deviatoric tensors included in the decomposition.
After this, it is proven, that these deviatoric tensors are independent,
  meaning they are no linear combination of one another.
In the end it is shown, that the decomposition is irreducible and orthogonal.
This work is concluded with various examples.
First, the recursive formula is used to find the irreducible and orthogonal
  deviatoric decomposition of general tensors up to order 4.
Then, with the help of calculations made by Hergl \cite{herglanalysis}, an exact calculation
  of the decomposition of the third order coupling tensor is given.
At last the calculation for the deviatoric decomposition for the fourth order stiffness tensor
  is demonstrated.
  
This work will not provide new facts about the deviatoric decomposition, but rather 
  concentrate on summarizing known results and explaining their derivation in more detail.

\section{Tensor Algebra}
An $N$-dimensional \textbf{tensor} of order $n$ $\mathds{T}^{(n)}$ can be defined 
  as a multilinear map of $n$ vectors to the real numbers
  \begin{equation*}
    \mathds{T}^{(n)} : (\mathcal{V}_N)^n \to \mathds{R}.
  \end{equation*}
In this work $N = 3$ will be assumed, unless   stated otherwise.
With a fixed orthonormal basis of $\mathcal{V}_N$, say $\{\textbf{e}_i\}$
    the tensor can be described by its coefficients by
  \begin{equation*}
    \mathds{T}^{(n)}(\textbf{e}_{i_1},\ldots,\textbf{e}_{i_n}) = T_{i_1 \cdots i_n}.
  \end{equation*}
This form will be referred to as the component form of a tensor.
Arbitrary $n$th order tensors will be denoted by $\mathds{T}^{(n)}$. 
  However zeroth order tensors, which can be represented by scalars,
  will be denoted by lower case greek letters, such as $\alpha$. 
First order tensors, which can be represented by vectors,
  will be denoted by bold lower case letters,
  such as $\textbf{v}$, unless stated otherwise.
  
For this work some tensor operations need
  to be defined.
One well known tensor operation is the tensor product or
  outer product. 
The tensor product of an $n$th-order tensor $\mathds{A}^{(n)}$ and 
  an $m$th-order tensor $\mathds{B}^{(m)}$ results in an $(n+m)$th-order tensor 
  $\mathds{C}^{(n+m)}$, defined by
\begin{equation*}
  C_{i_1 \dots i_n j_1 \dots j_m} = \mathds{A}^{(n)}\otimes \mathds{B}^{(m)} = A_{i_1 \dots i_n} B_{j_1 \dots j_m}.
\end{equation*}

Another well known tensor operation is called tensor contraction. 
The tensor contraction is the summation over a determined number 
  of indices. 
The single contraction $ \mathds{C}^{(n+m-2)} = \mathds{A}^{(n)} \cdot \mathds{B}^{(m)}$ is 
  the summation of two tensors $\mathds{A}^{(n)}$ and $\mathds{B}^{(m)}$ over
  one index
\begin{equation*}
  \begin{split}
     C_{i_1 i_2 \dots i_{n-1} j_2 \dots j_m } &= \mathds{A}^{(n)} \cdot \mathds{B}^{(m)} = \sum\limits_{k=1}^{N} A_{i_1 i_2 \dots i_{n-1} k} B_{k j_2 \dots j_m}. 
  \end{split} 
\end{equation*}

To reduce the overhead in the tensor formulas, the Einstein sum 
  convention can be used.
Here, an implicit summation over a repeated index is assumed:
\begin{equation*}
    \mathds{A}^{(n)} \cdot \mathds{B}^{(m)} = A_{i_1 i_2 \dots i_{n-1} k} B_{k j_2 \dots j_m}.
\end{equation*}
The double contraction $\mathds{A}^{(n)}: \mathds{B}^{(n)}$ is 
  analogously defined as the summation over two indices
\begin{equation*}
\begin{split}
C_{i_1 i_2 \dots i_{n-2} j_3 \dots j_m } &= \mathds{A}^{(n)} : \mathds{B}^{(m)} = \sum\limits_{k=1}^{N} \sum\limits_{l=1}^{N} A_{i_1 i_2 \dots i_{n-2} k l} B_{kl j_3 \dots j_m} \\
&= A_{i_1 i_2 \dots i_{n-2} kl} B_{kl j_3 \dots j_m}.
\end{split}
\end{equation*}

The complete contraction of the $n$th-order tensor $\mathds{A}^{(n)}$ and 
  the $m$th-order tensor $\mathds{B}^{(m)}$ with $m<n$, both of dimension $N$ 
  is used in this work by the following convention 
\begin{equation*}
\begin{split}
    \mathds{A}^{(n)}[\mathds{B}^{(m)}] &= A_{i_1 \dots i_m i_{m+1 \cdots i_n}} \left[B_{i_1\dots i_m} \right] \\
    &= \sum\limits_{i_1=1}^N \dots \sum\limits_{i_m=1}^N A_{i_1 \dots i_m i_{m+1} \cdots i_n} B_{i_1\dots  i_{m}}.
\end{split}
\end{equation*}
  
The main subject of this work is a tensor decomposition into irreducible, 
  orthogonal parts.
Orthogonal means that the rotation of the orthogonal parts
  results in the same rotation of the complete tensor and irreducible that the parts
  can not be decomposed further into rotational invariant parts.

Each of these parts can be represented by so called deviators which are
  totally symmetric and traceless tensors.
The subspace $\mathcal{D}^{(n)} \in
  (\mathcal{V}_N)^n$ denotes the space of all deviatoric tensors of order $n$.
A tensor is totally symmetric if it is invariant over all index permutations.
Each tensor can be symmetrized.
The symmetrization of a general tensor $\mathds{T}^{(n)}$ results in its 
  totally symmetric part $\textbf{s}\mathds{T}^{(n)}$ and is given by
  \begin{equation*}
    \textbf{s}\mathds{T}^{(n)} = \frac{1}{n!} \sum\limits_{\pi \in 
    \mathcal{S}_n} \pi \mathds{T}^{(n)}
  \end{equation*}
  where $\mathcal{S}_n$ is the set over all index permutations.

Another concept used in this work is the  
  symmetrization of an $n$th order tensor $\mathds{T}^{(n)}$ over a
  certain part of the index, for example $(i_1, \cdots, i_m), m<n$, 
  denoted by $T_{\hat{i_1} \cdots \hat{i_m} i_{m+1} \cdots i_n}$, as in \cite{zou2001orthogonal}. 
It is defined as the summation of all $(m!)$ permutations of 
  $T_{i_1 \cdots i_m i_{m+1} \cdots i_n)}$ on $(i_1 \cdots i_m)$ 
  divided by $m!$ 
\begin{align*}
    T_{\hat{i_1} \cdots \hat{i_m} i_{m+1} \cdots i_n}
    =
    \frac{1}{m!} \sum_{\pi \in \mathcal{S}_m} \pi \mathds{T}^{(n)}.
\end{align*}
If not the symmetrization but only a permutation of a certain group of indices is meant, 
  the notation $\pi(i_1, \cdots, i_n)$ will be used, where $\pi \in \mathcal{S}_n$.
The trace of an arbitrary $(n)$th-order tensor $\mathds{T}^{(n)}$ is dependent on the choice of a 
  specific index pair. 
For example the $(1,2)$-trace (trace over the first and second index) of $\mathds{T}^{(n)}$ is defined as an $(n-2)$th order tensor $\text{tr}_{1,2}(\mathds{T}^{(n)})$ as follows:
\begin{align*}
\text{tr}_{1,2}(\mathds{T}^{(n)}) := \sum_{s=1}^N T^{(n)}_{ssi_3 \cdots i_n}
\end{align*} 
In the totally symmetric case all traces of a tensor $\mathds{S}^{(n)}$ are the same \cite{hamermesh2012group}. 
Thus, in this work the trace of a totally symmetric tensor is defined as the $(1,2)$-trace.
\begin{equation*}
    \text{tr } ( \mathds{S} ) = \text{tr}_{1,2} ( \mathds{S} ) = \sum\limits_{s=1}^N S_{ss i_3 i_4 \dots i_n}.
  \end{equation*}
With this definition of the trace, a tensor is traceless, 
  if all traces (in the totally symmetric case this one trace) 
  vanish.
  
In the definition of the irreducible, orthogonal parts  hemitropic tensors,
  i.e. tensors that are invariant relative to the orthogonal group, are used.
For their definition two further very important tensors are used.
These two tensors are the second order identity tensor $\mathds{1}$ represented by the Kronecker delta 
\begin{equation*}
\delta_{ij} = \begin{cases}
1 & \text{ if } i=j,\\
0 & \text{ if } i\neq j
\end{cases}
\end{equation*}
  and the third order permutation tensor $\boldsymbol{\varepsilon}$
\begin{equation*}
    \varepsilon_{ijk} = \begin{cases}
    1 & \text{ if } (i,j,k) \in \left[ (1,2,3),\, (2,3,1),\, (3,1,2) \right], \\
    -1 & \text{ if } (i,j,k) \in \left[ (1,3,2),\, (2,1,3),\, (3,2,1) \right],\\
    0 & \text{else}.
    \end{cases}
\end{equation*}

Zou et al.\cite{zou2001orthogonal} imply that "every hemitropic tensor is expressible as a   linear combination of tensors in the form
\begin{align}
    &\text{per}(\mathds{1} \otimes \cdots \otimes \mathds{1}) \label{hemitropic_even}\\
    &\text{per}(\boldsymbol{\epsilon} \otimes \mathds{1} \otimes \cdots \otimes \mathds{1})\label{hemitropic_uneven}
\end{align}
where $\text{per}$ is a permutation operation."
(\ref{hemitropic_even}) is used for hemitropic tensors of even order, where as (\ref{hemitropic_uneven}) is used for hemitropic tensors of uneven order.
In this work $\mathds{H}^{(n)}$ will be used to denote 
  $n$th order hemitropic tensors.

\section{Deviatoric Decomposition of 
  Tensors in $\mathcal{V} \otimes \mathcal{D}^{(n)}$}
It will later be seen that the deviatoric decomposition
  for any $n$th order tensor $\mathds{T}^{(n)}$
  takes the form
\begin{align}\label{PreliminaryDecomp}
    \mathds{T}^{(n)} 
    = \sum_{J=1}^{J_0^n} \alpha_{J}\mathds{H}_J^{(n)}     
    +
    \sum_{J=1}^{J_1^n} \mathds{H}_J^{(n+1)} [\mathbf{v}_{J}] 
    + 
    \sum_{s=2}^{n} \sum_{J=1}^{J_s^n} \mathds{H}_J^{(n+s)}     [\mathds{D}_{J}^{(s)}]
\end{align}
where $J_s^n, 0 \le s \le n,$ represents the number of independent $s$th order deviatoric tensors. 

In order to prove the existence of the 
  deviatoric decomposition for any tensor $\mathds{T}^{(n)}$ of arbitrary order $n$, 
  further knowledge of a special tensor type is required. 
These special tensors $\mathds{G}^{(n+1)}$ belong to the space $\mathcal{V} \otimes \mathcal{D}^{(n)}$.

Regarding to Zou et al. \cite{zou2001orthogonal} for any $\textbf{v} = \sum_{k=1}^3 a_k \textbf{e}_k \in \mathcal{V}$, 
  $a_k \in \mathbb{R}$ yields:
  \begin{align*}
   \mathds{G}^{(n+1)} 
     = \textbf{v} \otimes \mathds{D}^{(n)} 
     = (\sum_{k = 1}^{3} a_k \textbf{e}_k) \otimes \mathds{D}^{(n)} 
     = \sum_{k = 1}^{3} (\textbf{e}_k \otimes  a_k \mathds{D}^{(n)})
     =:\sum_{k = 1}^{3} (\textbf{e}_k \otimes   \mathds{D}_k^{(n)})
  \end{align*}
Defining $\mathds{D}_k^{(n)} := a_k \mathds{D}^{(n)}$, 
  or $D^k_{i_1 \cdots i_n} = a_k D_{i_1 \cdots i_n}$,
  shows that the components of $\mathds{G}^{(n+1)}$ can be
  calculated as 
  \begin{align}\label{KomponentenG=KomponentenD}
  G_{k i_1 \cdots i_n} = D^k_{i_1 \cdots i_n}.    
  \end{align}
Therefore, $\mathds{G}^{(n+1)}$ is completely symmetric and traceless regarding
  the indices $(i_1 \cdots i_n)$.

Zheng et al. \cite{zheng2000irreducible} showed in detail 
  that the irreducible decomposition (\ref{PreliminaryDecomp}) of 
  $\mathds{G}^{(n)}$ for any integer $n > 1$ contains
  only the three deviatoric tensors $\mathds{D}^{(n-1)}$, $\mathds{D}^{(n)}$ 
  and $\mathds{D}^{(n+1)}$.
Therefore, it remains to specify the associated hemitropic tensors $\mathds{H}^{(2n)}$, 
  $\mathds{H}^{(2n+1)}$ and $\mathds{H}^{(2n+2)}$ respectively.
In the next part each irreducible part will be analyzed separately. 

\textbf{Analysis of $\mathbf{\mathds{H}^{(2n)}[\mathds{D}^{(n-1)}]}$}

The hemitropic tensor needs to be of order $2n$, because
  the complete contraction with the $(n-1)$th order deviatoric tensor 
  must be of order $(n+1)$.
Since $2n$ is even for any $n$, $\mathds{H}^{(2n)}$ must be a
  linear combination of $\text{per}(\mathds{1} \otimes ... \otimes \mathds{1})$,
  or $  \delta_{\pi(i_1 j_1} \delta_{i_2 j_2} ... 
  \delta_{i_{n-1} j_{n-1}} \delta_{i_n k)} $, see (\ref{hemitropic_even}).
The different permutations of $(i_1 j_1 i_2 j_2 \cdots i_{n-1} j_{n-1} i_n k) $ 
  affect the indices of the various $\boldsymbol{\delta}$'s, therefore, it is 
  important to examine each summand of the linear combination closely.

Let $x_1, \cdots ,x_{n-1} \in \{1, \cdots ,n-1\}$ and $y_1, \cdots , y_n \in \{1,...,n\}$.
If a summand of the linear combination contains 
  at least one $\delta_{j_{x_1} j_{x_2}}$, 
  the whole term is equal to zero, because $\mathds{D}^{(n-1)}$ is deviatoric, thus, traceless
  and therefore, 
\begin{equation} \nonumber
    \delta_{j_{x_1} j_{x_2}}D_{j_{x_1} j_{x_2} \pi(j_3 \cdots j_{n-1})} = 0
\end{equation}
Suppose a summand does not contain any $\delta_{j_{x_1} j_{x_2}}$, then for the other possibilities yields:

\begin{enumerate}
    \item[\textbf{Case 1:}] 
    The summand is given by $\delta_{\pi ( i_{y_n}j_{x_1}} \delta_{i_{y_2}j_{x_2}} \dots \delta_{i_{y_{n-1}}j_{x_{n-1})}}\delta_{i_{y_1}k} $.
    Then, the tensor contraction with $\mathds{D}^{(n-1)}$ is zero except for the case 
    $\delta_{i_{y_n}j_{x_1}} \delta_{i_{y_2}j_{x_2}} \dots \delta_{i_{y_{n-1}}j_{x_{n-1}}}\delta_{i_{y_1}k} $.
    $\delta_{i_{y_a}j_{x_a}} $ 
    with $i_{y_1} = k$, $i_{y_n} = j_{x_1}$, $i_{y_2} = j_{x_2} \cdots, i_{y_{n-1}} = j_{x_{n-1}}$. 
    The tensor contraction can then be simplified to the following multiplication:
    \begin{align*}
     \delta_{k i_{y_1}} D_{i_{y_2} \cdots i_{y_n}}.
    \end{align*}

    \item[\textbf{Case 2:}] 
    The summand is given by $\delta_{\pi ( i_{y_2}j_{x_2}} \dots \delta_{i_{y_{n-1}}j_{x_{n-1}}} \delta_{i_{y_{1}}i_{y_{n})}}\delta_{j_{x_1}k} $.
    Then, the tensor contraction with $\mathds{D}^{(n-1)}$ is zero except for the case
    $\delta_{i_{y_n}j_{x_2}} \delta_{i_{y_3}j_{x_3}} \dots \delta_{i_{y_{1}}i_{y_{2}}}\delta_{j_{x_1}k} $.
        $\delta_{i_{y_a}j_{x_a}} $ 
    with $j_{x_1} = k$, $i_{y_1} = i_{y_2}$, $i_{y_n} = j_{x_2}$, $i_{y_3} = j_{x_3}
    \cdots, i_{y_{n-1}} = j_{x_{n-1}}$. 
    The tensor contraction can then be simplified to the following multiplication:
    \begin{align*}
     \delta_{i_{y_1} i_{y_2}} D_{i_{y_3} \cdots i_{y_n} k}. 
    \end{align*}
\end{enumerate}

In result the contraction contains only the $(n+1)!$
  summands $\delta_{\pi(i_1 i_2} D_{i_3 \cdots i_nk)}$, 
  with $\pi \in S_{(n+1)}$. 
Therefore, it is expressible as a linear combination of $\text{per}\left(\mathds{1} \otimes \mathds{D}^{(n-1)}\right)$.

$\mathds{1}$ and $\mathds{D}^{(n-1)}$ are completely symmetric.
Thus, there are two types of summands: two summands 
  have the form $\delta_{k \pi(i_1} D_{i_2 \cdots i_{n})}$ and 
  $(n-1)$ summands have the form $\delta_{\pi(i_1 i_2} D_{i_3 \cdots i_n) k}$, 
  where $\pi \in S_n$. 

This can be summarized as follows:
\begin{align*}
    H_{i_1 \cdots i_nkj_1 \cdots j_{n-1}}[D_{j_1 \cdots j_{n-1}}]
    &=
   (\sum_{\lambda = 1, \pi \in S_{2n}}^{(2n)!}  
     c_{\lambda}  \delta_{\pi(i_1 j_1} \delta_{i_2 j_2} \cdots 
     \delta_{i_{n-1} j_{n-1}} \delta_{i_n k)} 
    ) 
     [D_{j_1 \cdots j_{n-1}}] \\
   &=
   \sum_{\lambda = 1, \pi \in S_{n+1}}^{(n+1)!}
   c_{\lambda} \delta_{\pi(i_1 i_2} D_{i_3 \cdots i_{n} k)} \\
   &=
   \sum_{\lambda = 1, \pi \in S_{n}}^{n!}
     \sum_{\mu = 1}^{2}  c_{\lambda_{\mu}}  \delta_{k \pi(i_1} D_{i_2 \cdots i_{n})} + 
   \sum_{\lambda = 1, \pi \in S_{n}}^{n!}
     \sum_{\mu = 3}^{n+1} c_{\lambda_{\mu}}  \delta_{\pi(i_1 i_2} D_{i_3 \cdots i_n) k}\\
    &=
   \sum_{\lambda = 1, \pi \in S_{n}}^{n!}
     a \delta_{k \pi(i_1} D_{i_2 \cdots i_{n})} + 
   \sum_{\lambda = 1, \pi \in S_{n}}^{n!}
     b \delta_{\pi(i_1 i_2} D_{i_3 \cdots i_n) k}
\end{align*}
where $a = \sum_{\mu = 1}^{2}  c_{\lambda_{\mu}} $ and 
  $b = \sum_{\mu = 3}^{n+1} c_{\lambda_{\mu}}$. 
One equation with two unknown 
  variables $a, b \in \mathbb{R}$ is gained.

\textbf{Analysis of $\mathbf{\mathds{H}^{(2n+1)}[\mathds{D}^{(n)}]}$}

The hemitropic tensor needs to be of order $(2n+1)$, because
  the complete contraction with the $n$th order deviatoric tensor 
  must be of order $(n+1)$.
Since $2n+1$ is odd for any $n$, $\mathds{H}^{(2n+1)}$ must be a
  linear combination  of 
  $\text{per}(\mathbf{\boldsymbol{\epsilon}} \otimes \mathds{1} \otimes \cdots \otimes \mathds{1})$,
  or $\epsilon_{\pi(i_1 j_1 k} \delta_{i_2 j_2} \cdots \delta_{i_n j_n)}$, see (\ref{hemitropic_uneven}).
Again, the different permutations of $(i_1 j_1 i_2 j_2 \cdots i_{n} j_{n} k) $ 
  affect the indices of the various $\boldsymbol{\delta}$'s and of $\boldsymbol{\epsilon}$, 
  therefore, it is 
  important to examine each summand of the linear combination closely.
  
Let $x_1, \cdots , x_n \in \{1, \cdots , n \}$ and
  $y_1, \cdots  , y_n \in \{1, \cdots , n\}$.
As before, if a summand of the linear combination contains at least one $\delta_{j_{x_1} j_{x_2}}$, the
  whole summand is equal to zero, because
  $\mathds{D}^{(n)}$ is deviatoric, thus, traceless and therefore
\begin{equation} \nonumber
    \delta_{j_{x_1} j_{x_2}}D_{j_{x_1} j_{x_2} \pi(j_3 \cdots j_{n})} = 0.
\end{equation}
Suppose a summand does not contain any $\delta_{j_{x_1} j_{x_2}}$. 
Then, other possible summands are given by:
      
\begin{enumerate}
    \item[\textbf{Case 1:}]
    The summand is given by 
    $\delta_{\pi ( i_{y_4}j_{x_4}} \dots 
    \delta_{i_{y_{n}}j_{x_{n}}} 
    \delta_{i_{y_{1}}i_{y_{2}}}
    \delta_{i_{y_{3} k})}
    \epsilon_{j_{x_1} j_{x_2} j_{x_3}}$.
    Independent of the following terms containing any $\delta$ the following tensor contraction of $\boldsymbol{\epsilon}$ and
      $\mathds{D}^{(n)}$ results in \\
      \begin{align*}
          \epsilon_{j_{x_1} j_{x_2} j_{x_3}} D_{j_{x_1} j_{x_2} j_{x_3} j_{x_4} ... j_{x_n}} 
          &=
          \epsilon_{1 1 1} D_{1 1 1 j_{x_4} \cdots j_{x_n}} + 
          \epsilon_{2 2 2} D_{2 2 2 j_{x_4} \cdots j_{x_n}} + 
          \epsilon_{3 3 3} D_{3 3 3 j_{x_4} \cdots j_{x_n}} \\ &+ 
          \epsilon_{1 2 3} D_{1 2 3 j_{x_4} \cdots j_{x_n}} + 
          \epsilon_{1 3 2} D_{1 2 3 j_{x_4} \cdots j_{x_n}} \\ &+ 
          \epsilon_{2 1 3} D_{2 1 3 j_{x_4} \cdots j_{x_n}} + 
          \epsilon_{2 3 1} D_{2 3 1 j_{x_4} \cdots j_{x_n}} \\ &+ 
          \epsilon_{3 1 2} D_{3 1 2 j_{x_4} \cdots j_{x_n}} + 
          \epsilon_{3 2 1} D_{3 2 1 j_{x_4} \cdots j_{x_n}} \\
          &= 0
      \end{align*}
   
    \item[\textbf{Case 2:}]
    The summand is given by 
    $\delta_{\pi ( i_{y_3}j_{x_3}} \dots 
    \delta_{i_{y_{n}}j_{x_{n}}} 
    \delta_{i_{y_{1}}i_{y_{2}}}
    \delta_{i_{y_{2} k})}
    \epsilon_{j_{x_1} j_{x_2} i_{y_1}}$.
    Independent of the following terms containing any $\delta$ the following tensor contraction of $\boldsymbol{\epsilon}$ and
      $\mathds{D}^{(n)}$ results in \\
      \begin{align*}
          \epsilon_{j_{x_1} j_{x_2} i_{y_1}} D_{j_{x_1} j_{x_2} i_{y_1} j_{x_4} ... j_{x_n}} 
          &=
          \epsilon_{1 1 i_{y_1}} D_{1 1 i_{y_1} j_{x_4} \cdots j_{x_n}} + 
          \epsilon_{2 2 i_{y_1}} D_{2 2 i_{y_1} j_{x_4} \cdots j_{x_n}} + 
          \epsilon_{3 3 i_{y_1}} D_{3 3 i_{y_1} j_{x_4} \cdots j_{x_n}} \\ &+ 
          \epsilon_{1 2 i_{y_1}} D_{1 2 i_{y_1} j_{x_4} \cdots j_{x_n}} + 
          \epsilon_{2 1 i_{y_1}} D_{2 1 i_{y_1} j_{x_4} \cdots j_{x_n}} \\ &+ 
          \epsilon_{1 3 i_{y_1}} D_{1 3 i_{y_1} j_{x_4} \cdots j_{x_n}} + 
          \epsilon_{3 1 i_{y_1}} D_{3 1 i_{y_1} j_{x_4} \cdots j_{x_n}} \\ &+ 
          \epsilon_{2 3 i_{y_1}} D_{2 3 i_{y_1} j_{x_4} \cdots j_{x_n}} + 
          \epsilon_{3 2 i_{y_1}} D_{3 2 i_{y_1} j_{x_4} \cdots j_{x_n}} \\
          &= 0
      \end{align*}
    
    \item[\textbf{Case 3:}]
    The summand is given by 
    $\delta_{\pi ( i_{y_3}j_{x_3}} \dots 
    \delta_{i_{y_{n}}j_{x_{n}}} 
    \delta_{i_{y_{1}}i_{y_{2})}}
    \epsilon_{j_{x_1} j_{x_2} k}$.
    Independent of the following terms containing any $\delta$ the following tensor contraction of $\boldsymbol{\epsilon}$ and
      $\mathds{D}^{(n)}$ results in \\
      \begin{align*}
          \epsilon_{j_{x_1} j_{x_2} k} D_{j_{x_1} j_{x_2} k j_{x_4} \cdots j_{x_n}} 
          &=
          \epsilon_{1 1 k} D_{1 1 k j_{x_4} \cdots j_{x_n}} + 
          \epsilon_{2 2 k} D_{2 2 k j_{x_4} \cdots j_{x_n}} + 
          \epsilon_{3 3 k} D_{3 3 k j_{x_4} \cdots j_{x_n}} \\ &+ 
          \epsilon_{1 2 k} D_{1 2 k j_{x_4} \cdots j_{x_n}} + 
          \epsilon_{2 1 k} D_{2 1 k j_{x_4} \cdots j_{x_n}} \\ &+ 
          \epsilon_{1 3 k} D_{1 3 k j_{x_4} \cdots j_{x_n}} + 
          \epsilon_{3 1 k} D_{3 1 k j_{x_4} \cdots j_{x_n}} \\ &+ 
          \epsilon_{2 3 k} D_{2 3 k j_{x_4} \cdots j_{x_n}} + 
          \epsilon_{3 2 k} D_{3 2k j_{x_4} \cdots j_{x_n}} \\
          &= 0
      \end{align*}

    \item[\textbf{Case 4:}]
    The summand is given by 
    $\delta_{\pi ( i_{y_3}j_{x_3}} \dots 
    \delta_{i_{y_{n}}j_{x_{n}}} 
    \delta_{j_{x_{2}} k)}
    \epsilon_{j_{x_1} i_{y_1} i_{y_2}}$.
    Then, the tensor contraction with $\mathds{D}^{(n)}$ is zero except for the case 
    $\epsilon_{j_{x_1}i_{y_1} i_{y_2}}\delta_{i_{y_3}j_{x_3}} \dots \delta_{i_{y_n}j_{x_n}} \delta_{j_{x_{2}} j_k} $.
    
    Where $\delta_{i_{y_a}j_{x_a}} $ 
    with $j_{x_2} = k$, $i_{y_2} = j_{x_2}, \cdots, i_{y_{n}} = j_{x_{n}}$ and
    $\epsilon_{j_{x_1} i_{y_1} i_{y_2}}$ with $j_{x_1} \neq i_{y_1} \neq i_{y_2}$. 
    The tensor contraction can then be simplified to a single contraction:
    \begin{align*}
     \epsilon_{j_{x_1} i_{y_1} i_{y_2}} D_{j_{x_1} k i_{y_3} \cdots i_{y_n}}
    \end{align*}

    \item[\textbf{Case 5:}]
    The summand is given by 
    $\delta_{\pi ( i_{y_2}j_{x_2}} \dots 
    \delta_{i_{y_{n}}j_{x_{n}})} 
    \epsilon_{j_{x_1} i_{y_1} k}$.
    Then, the tensor contraction with $\mathds{D}^{(n)}$ is zero except for the case 
    $\epsilon_{j_{x_1}i_{y_1}k}\delta_{i_{y_2}j_{x_2}} \dots \delta_{i_{y_n}j_{x_n}} $.
    
    Where $\delta_{i_{y_a}j_{x_a}} $ 
    with $i_{y_1} = j_{x_1}, \cdots, i_{y_{n}} = j_{x_{n}}$ and $\epsilon_{j_{x_1} i_{y_1} k}$
    with $j_{x_1} \neq i_{y_1} \neq k$. 
    The tensor contraction can then be simplified to a single contraction:
    \begin{align*}
     \epsilon_{j_{x_1} i_{y_1} k} D_{j_{x_1} i_{y_2} \cdots i_{y_n}}
    \end{align*}
\end{enumerate}

The final linear combination will contain all summands of the form
  $\epsilon_{j_{x_1} i_{y_1} i_{y_2}} D_{j_{x_1} k i_{y_3} \cdots i_{y_n}}$
  and  $\epsilon_{j_{x_1} i_{y_1} k} D_{j_{x_1} i_{y_2} \cdots i_{y_n}}$.
Therefore, it is expressible as a linear combination of
 permutations of $(\boldsymbol{\epsilon} \cdot \mathds{D}^{(n-1)})$.

As a next step the position of the index variable $k$ and the index variable$j_{x_a}$, 
  which is used for the contraction,
  will be examined closer.

\begin{itemize}
    \item The linear combination will include all summands in which only
    the index, that is used for the contraction $j_{x_a}$, varies. 
    Note that due to the definition of a contraction each 
      contraction index $j_{i_1}, \cdots, j_{i_n}$ will result in the same term.
    Therefore the summand is independent of the choice of $j_{i_1}, \cdots, j_{i_n}$ and
      in the following $s$ will be used to mark the contraction index.

    \item The position of the contraction variable $s$ within $\mathds{D}^{(n-1)}$ results in the same term and, therefore, may have a set position. 
    
    \item The position of the contraction variable $s$ within $\boldsymbol{\epsilon}$ results in the same term, except for a multiplication with $1$ or $-1$. 
    Therefore, it may also have a set position. 
    The sign change will be noted in the scalar of the linear combination.
\end{itemize}

These facts can be summarized as follows:
\begin{align}
      H_{i_1 \cdots i_n k j_1 \cdots j_n} D_{j_1 \cdots j_n} \nonumber
     &=
     \sum_{\lambda = 1, \pi \in S_{2n+1}}^{(2n+1)!} 
      c_\lambda \epsilon_{\pi(i_1 j_1 k} \delta_{i_2 j_2} \cdots \delta_{i_n j_n)} D_{j_1 \cdots j_n} \\ 
     &=
     \sum_{\lambda = 1, \pi \in S_{n}}^{n!}
      d \epsilon_{s \pi(i_1 i_2} D_{i_3 \cdots i_n) k s} +
      e \epsilon_{k s \pi(i_1} D_{i_2 \cdots i_n) s} \label{d_zero_duetosymmetry}\\
     &=
     \sum_{\lambda = 1, \pi \in S_{n}}^{n!}
      e \epsilon_{k s \pi(i_1} D_{i_2 \cdots i_n) s} \nonumber .
\end{align}
Evaluating line (\ref{d_zero_duetosymmetry})
causes all summands of the form $d \epsilon_{s \pi(i_1 i_2} D_{i_3 \cdots i_n) k s}$
  to vanish, due to the skew symmetry of $\boldsymbol{\epsilon}$.
Therefore one equation with one unknown 
  variable $e \in \mathbb{R}$ is gained. \\ \\ \\

\textbf{Analysis of $\mathbf{\mathds{H}^{(2n+2)}[\mathds{D}^{(n+1)}]}$}

The hemitropic tensor needs to be of order $(2n+2)$, because
  the complete contraction with the $(n+1)$th order deviatoric tensor 
  must be of order $(n+1)$.
  Since $2n+2$ is even for any $n$, $\mathds{H}^{(2n+2)}$ must be a
  linear combination of $\text{per}(\mathds{1} \otimes ... \otimes \mathds{1})$,
  or $  \delta_{\pi(i_1 j_1} \delta_{i_2 j_2} ... 
  \delta_{i_{n} j_{n}} \delta_{i_{n+1} k)} $, see (\ref{hemitropic_even}).
  
Let $x_1, \cdots ,x_{n+1} \in \{1, \cdots ,n+1\}$ and $y_1, \cdots , y_n \in \{1,...,n\}$.
If the summand contains 
  at least one $\delta_{j_{x_1} j_{x_2}}$ or $\delta_{j_{x_1} k}$, 
  the whole term is equal to zero, because $\mathds{D}^{(n+1)}$ is deviatoric, thus, traceless
  and therefore, 
\begin{equation} \nonumber
    \delta_{j_{x_1} j_{x_2}}D_{j_{x_1} j_{x_2} \pi(j_3 \cdots j_{n}) k} = 0
\end{equation} or
\begin{equation} \nonumber
    \delta_{j_{x_1} k} D_{j_{x_1} k \pi(j_2 \cdots j_{n}) } = 0.
\end{equation}
Suppose a summand does not contain any $\delta_{j_{x_1} j_{x_2}}$, or $\delta_{j_{x_1} k}$. 
Then for the other possibilities yields
that the summand can only be given by
\begin{equation}
    \delta_{j_{x_1} i_{y_1}} \dots  \delta_{j_{x_n} i_{y_n}}  \delta_{k i_{y_{n+1}}}. \nonumber
\end{equation}
Then, the tensor contraction with $\mathds{D}^{(n+1)}$ is zero except for the case where $j_{x_1} =   i_{y_1}, \dots, j_{x_n} = i_{y_n}, k = i_{x_{n+1}} $. 
Due to the symmetrie of $\mathds{D}^{(n+1)}$ the final tensor contraction takes the following form:
\begin{equation} \nonumber
    H_{k i_1 \dots i_{n+1} j_1 \dots j_n} D_{k j_1 \dots j_n} 
    =
    \sum_{\lambda = 1}^{(n+1)!} c_{\lambda} D_{k i_1 \dots i_n} 
    = f D_{k i_1 \dots i_n}.
\end{equation}

\textbf{Trace Calculation}

With all the previous facts the following equation for $\mathds{G}^{(n+1)}$ was gained:
\begin{equation} \label{GgemischteSkalare}
    G_{k i_1 \dots i_n} = 
    \sum_{\lambda = 1, \pi \in S_{n}}^{n!}
    a\delta_{k \pi(i_1} D_{i_2 \cdots i_{n})} 
   + 
    \sum_{\lambda = 1, \pi \in S_{n}}^{n!}
   b  \delta_{\pi(i_1 i_2} D_{i_3 \cdots i_n) k}  
   +
   \sum_{\lambda = 1, \pi \in S_{n}}^{n!}
   d \epsilon_{k s \pi(i_1} D_{i_2 \cdots i_n) s}
   + f D_{k i_1 \dots i_n}
\end{equation}

From (\ref{KomponentenG=KomponentenD}) it is already known that $\mathds{G}^{(n+1)}$
  is completely symmetric and traceless regarding the indices $(i_1 \cdots i_n)$.

The representation (\ref{GgemischteSkalare}) is, due to the symmetrization, completely symmetric regarding
  $(i_1 \cdots i_n)$. 
The fact, that it is also traceless can be used to find solutions 
  for $a$, $b$, $d$ and $f$.
It follows
\begin{align*}
    0 = 
    \delta_{i_1 i_2} [ 
     \sum_{\lambda = 1, \pi \in S_{n}}^{n!}
   a \delta_{k \pi(i_1} D_{i_2 \cdots i_{n})} + 
     \sum_{\lambda = 1, \pi \in S_{n}}^{n!}
   b  \delta_{\pi(i_1 i_2} D_{i_3 \cdots i_n) k}  
   +
   \sum_{\lambda = 1, \pi \in S_{n}}^{n!}
   d \epsilon_{k s \pi(i_1} D_{i_2 \cdots i_n) s}
   + f D_{k i_1 \dots i_n}] .
   \end{align*}

Due to $\mathds{D}^{(n)}$ and $\mathds{D}^{(n+1)}$ being traceless and  
  $\boldsymbol{\epsilon}$ being skew-symmetric it yields
\begin{align*}
    0 = 
    \delta_{i_1 i_2} [ 
   f D_{k i_1 \dots i_n}] 
\end{align*}
and
\begin{align*}
    0 = 
    \delta_{i_1 i_2} [ 
   \sum_{\lambda = 1, \pi \in S_{n}}^{n!}
   d \epsilon_{k s \pi(i_1} D_{i_2 \cdots i_n) s}].
\end{align*}
Therefore, $d$ and $f$ may be freely chosen as $d=\frac{1}{n!}$ and $e = 1$.

It remains to analyze the condition for
\begin{align*}
    0 = 
    \delta_{i_1 i_2} [ 
     \sum_{\lambda = 1, \pi \in S_{n}}^{n!}
   a \delta_{k \pi(i_1} D_{i_2 \cdots i_{n})} + 
     \sum_{\lambda = 1, \pi \in S_{n}}^{n!}
   b  \delta_{\pi(i_1 i_2} D_{i_3 \cdots i_n) k}].
\end{align*}

The key to solve this equation is to look at all possibilities where $i_1$ and $i_2$ can
  be located within the different summands.

These will be analyzed in the following:

\begin{enumerate}
    \item[\textbf{Case 1:}] The summand is given by $\delta_{i_1i_2}D_{\pi (i_3 \dots i_n)k}$.
    Then it is
    \begin{align*}
                \delta_{i_1 i_2} [\delta_{i_1 i_2} D_{\pi(i_3 \cdots i_n) k}] 
                &= \delta_{1 1} D_{\pi(i_3 \cdots i_n) k} + \delta_{2 2} D_{\pi(i_3 \cdots i_n) k} 
                  + \delta_{3 3} D_{\pi(i_3 \cdots i_n) k} \\
                &= 3D_{\pi(i_3 \cdots i_n) k}.
            \end{align*}
            
    There are zero summands of this kind for $\sum_{\lambda = 1, \pi \in S_{n}}^{n!}
   a \delta_{k \pi(i_1} D_{i_2 \cdots i_{n})}$.
   
    There are two summands of this kind for $\sum_{\lambda = 1, \pi \in S_{n}}^{n!}
   b  \delta_{\pi(i_1 i_2} D_{i_3 \cdots i_n) k}$.
    
    \item[\textbf{Case 2:}] The summand is given by $\delta_{\pi (ki_3} D_{i_4 \dots i_n)i_1i_2}$. Then it is
    \begin{align*}
                \delta_{i_1 i_2} [\delta_{\pi (k i_3} D_{i_4 \cdots i_n) i_1 i_2}] 
                &= \delta_{\pi (k i_3}\text{tr}_{n-1,n} \left( D_{i_4 \dots i_n) i_1 i_2} \right) = 0
    \end{align*}
    because $\mathds{D}^{(n-1)}$ is traceless.

    There are $(n-1)(n-2)$ summands of this kind for $\sum_{\lambda = 1, \pi \in S_{n}}^{n!}
   a \delta_{k \pi(i_1} D_{i_2 \cdots i_{n})}$.
   
    There are $(n-2)(n-3)$ summands of this kind for $\sum_{\lambda = 1, \pi \in S_{n}}^{n!}
   b  \delta_{\pi(i_1 i_2} D_{i_3 \cdots i_n) k}$.
   
    \item[\textbf{Case 3:}] The summand is given by $\delta_{i_1 \pi (k}D_{i_3 \dots i_n )i_2}$. Then it is
        \begin{align*}
                \delta_{i_1 i_2} [\delta_{k i_1} D_{i_2 \pi(i_3 \cdots i_{n})}]
                &= \delta_{k 1} D_{1 \pi(i_3 \cdots i_{n}} + \delta_{k 2} D_{2 \pi(i_3 \cdots i_{n})} 
                   + \delta_{k 3} D_{3 \pi(i_3 \cdots i_{n})} \\
                &= D_{k \pi(i_3 \cdots i_{n})}
        \end{align*}
        There are
        and 
        \begin{align*}
                \delta_{i_1 i_2} [\delta_{\pi(i_n i_1} D_{i_2 \cdots i_{n-1}) k}]
                &= \delta_{\pi(i_n 1} D_{1 \cdots i_{n-1}) k} + 
                   \delta_{\pi(i_n 2} D_{2 \cdots i_{n-1}) k} 
                   + \delta_{\pi(i_n 3} D_{3 \cdots i_{n-1}) k} \\
                &= D_{\pi(i_n i_3 \cdots i_{n-1}) k}. 
        \end{align*}

    There are $2(n-1)$ summands of this kind for $\sum_{\lambda = 1, \pi \in S_{n}}^{n!}
   a \delta_{k \pi(i_1} D_{i_2 \cdots i_{n})}$.
   
    There are $4(n-2)$ summands of this kind for $\sum_{\lambda = 1, \pi \in S_{n}}^{n!}
   b  \delta_{\pi(i_1 i_2} D_{i_3 \cdots i_n) k}$.
\end{enumerate}

All these facts put together, the following equation for the trace is gained: 
\begin{align}
    0 \nonumber
    &= 
    \delta_{i_1 i_2} [ 
     \sum_{\lambda = 1, \pi \in S_{n}}^{n!}
       a  \delta_{k \pi(i_1} D_{i_2 \cdots i_{n})} + 
     \sum_{\lambda = 1, \pi \in S_{n}}^{n!}
       b  \delta_{\pi(i_1 i_2} D_{i_3 \cdots i_n) k}]  \\
    &= \nonumber
    \sum_{\lambda = 1, \pi \in S_{n-2}}^{(n-2)!}
      (2(n-1)aD_{\pi(i_3 \cdots i_n) k}) 
      + 
    \sum_{\lambda = 1, \pi \in S_{n-2}}^{(n-2)!}
      b(6D_{\pi(i_3 \cdots i_n) k} + 4(n-2)D_{\pi(i_3 \cdots i_n) k}) \\
    &= \nonumber
    \sum_{\lambda = 1, \pi \in S_{n-2}}^{(n-2)!}
    ((2n-2)a + (4n-2)b)D_{\pi(i_3 \cdots i_n) k}\\
    &= \label{trace_a_b}
    D_{i_3 \cdots i_n k} 
    \sum_{\lambda = 1}^{(n-2)!}
    (2n-2)a + (4n-2)b
    \end{align}

If $D_{i_3...i_n k} \neq 0$ and $a \neq 0 \neq b$ is assumed,
  line (\ref{trace_a_b}) is true 
  for
    \begin{align}\label{ab}
    \frac{n-1}{2n-1} a &= - b
    \end{align}

Therefore, one possible solution is given by
  $a = \frac{2n-1}{n!(n-1)}$ and $b = \frac{-1}{n!}$.

All together the component form for $\mathds{G}^{(n+1)}$ is gained as described by Zou et al. \cite{zou2001orthogonal}:
\begin{equation}\label{finalG}
    G_{k i_1 \dots i_n} 
    = 
    \frac{2n-1}{n-1} \delta_{k \hat{i}_1} D_{\hat{i}_2 \dots \hat{i}_n}
    - \delta_{\hat{i}_1 \hat{i}_2} D_{\hat{i}_3 \dots \hat{i}_n k}
    + \epsilon_{ks\hat{i}_1}D_{\hat{i}_2 \dots \hat{i}_n s}
    + D_{k i_1 \dots i_n}
\end{equation}

In line (\ref{ab}) it can be seen that
  (\ref{finalG}) is not unique up to a scalar multiplication. \\ \\ \\ \\ \\ \\

\section{Prove of the Existence of the Deviatoric Decomposition}
The main goal of this work is to prove that any tensor of arbitrary order in 3 dimensions can be decomposed into an irreducible and orthogonal sum of deviatoric tensors:
\begin{align} \label{ZerlegungT(n)}
    \mathds{T}^{(n)} 
    = \sum_{J=1}^{J_0^n} \alpha_{J}\mathds{H}_J^{(n)}     
    +
    \sum_{J=1}^{J_1^n} \mathds{H}_J^{(n+1)} [\mathbf{v}_{J}] 
    + 
    \sum_{s=2}^{n} \sum_{J=1}^{J_s^n} \mathds{H}_J^{(n+s)}     [\mathds{D}_{J}^{(s)}]
\end{align}
or equivalently 
\begin{align} \label{ZerlegungT(n)Komponenten}
    T_{i_1 \cdots i_n}
    =
    \sum_{J=1}^{J_0^n} \alpha_J H_{i_1 \cdots i_n}^{J}
    +
    \sum_{J=1}^{J_1^n} H_{i_1 \cdots i_n j}^{J}[v_j^J] 
    + 
    \sum_{s=2}^{J_s^n} 
    \sum_{J=1}^{J_0^n} H_{i_1 \cdots i_n j_1 \cdots j_s}^{J}[D_{j_1 \cdots j_n}^{J}]
\end{align}
where $J_s^n, 0 \le s \le n,$ represents the number of independent $s$th order deviatoric tensors. 

This statement can be proven by induction.
It is well known that any second order tensor $\mathds{T}^{(2)}$ can 
  "be decomposed into the sum of its symmetric and anti symmetric part, and the 
  symmetric part can further be decomposed into the sum of deviatoric and
  hydro static part" \cite{zou2001orthogonal}. 
Therefore, a second order tensor takes the following form:
\begin{align}\label{decomp2order}
    \mathds{T}^{(2)} 
    = 
    \alpha \mathbf{1} 
    + 
    \mathbf{\epsilon} \cdot \textbf{v}
    +
    \mathds{D}^{(n)}
\end{align}
This concludes the base case.
For the inductive step, assume that (\ref{ZerlegungT(n)}) is true for 
  a certain integer $n$. Now this statement needs to be proven for $(n+1)$. 

In general any $(n+1)$th order tensor can be written as the tensor product between an arbitrary vector $\mathbf{v} = c_k \textbf{e}_k \in \mathcal{V}, c_k \in \mathds{R}$ and an $n$th order tensor $\mathds{T}^{(n)}$ by
\begin{align*}
    \mathds{T}^{(n+1)} 
    =
    \mathbf{v} \otimes \mathds{T}^{(n)}
    = 
    (c_k \textbf{e}_k) \otimes \mathds{T}^{(n)}
    =
    \textbf{e}_k \otimes c_k \mathds{T}^{(n)}
    =:
    \textbf{e}_k \otimes \mathds{T}_k^{(n)}.
\end{align*}

Therefore, there are two representations of $\mathds{T}^{(n+1)}$, which need to be identified and then checked, if they are equal:
\begin{enumerate}
    \item  $\textbf{e}_k \otimes \mathds{T}_k^{(n)}$ is calculated by using (\ref{ZerlegungT(n)}) (representation 1),
    \item  (\ref{ZerlegungT(n)}) is used with $(n+1)$ instead of $(n)$ (representation 2).
\end{enumerate}

\textbf{Representation 1}

Suppose an $(n+1)$th order tensor can be represented as $\mathds{T}^{(n+1)} = \mathbf{e}_k \otimes \mathds{T}_k^{(n)}$. Assuming the deviatoric decomposition for $\mathds{T}_k^{(n)}$ exists, the following first result for $\mathds{T}^{(n+1)}$ is established. It is given by
\begin{align*}
\mathds{T}^{(n+1)}
= 
\mathbf{e}_k \otimes \mathds{T}_k^{(n)} 
= \mathbf{e}_1 \otimes \mathds{T}_1^{(n)} + \mathbf{e}_2 \otimes \mathds{T}_2^{(n)} + \mathbf{e}_3 \otimes \mathds{T}_3^{(n)}.
\end{align*} 

Using the deviatoric decomposition for $\mathds{T}_k$ will result in a large formula.
Thus, three dots are used as placeholders for the irreducible parts. In the following calculation certain summands will be examined separately before putting them back together in the end.
The deviatoric decompositon of $\mathds{T}^{(n+1)} $ is calculated by
\begin{align*}
\mathds{T}^{(n+1)}
&=
\mathbf{e}_1 \otimes (\sum_{J=1}^{J_0^n} (\cdots)
+
\sum_{J=1}^{J_1^n} (\cdots)
+ 
\sum_{s=2}^{n} \sum_{J=1}^{J_s^n} (\cdots))\\
&+ 
\mathbf{e}_2 \otimes (\sum_{J=1}^{J_0^n} (\cdots)
+
\sum_{J=1}^{J_1^n} (\cdots)
+ 
\sum_{s=2}^{n} \sum_{J=1}^{J_s^n} (\cdots))  \\
&+ 
\mathbf{e}_3 \otimes (\sum_{J=1}^{J_0^n} (\cdots)
+
\sum_{J=1}^{J_1^n} (\cdots)
+ 
\sum_{s=2}^{n} \sum_{J=1}^{J_s^n} (\cdots)) \\
&= \underbrace{\mathbf{e}_1 \otimes (\sum_{J=1}^{J_0^n} (\cdots))}_{(a_1)}
+
\underbrace{\mathbf{e}_1 \otimes\sum_{J=1}^{J_1^n} (\cdots))}_{(b_1)}
+ 
\underbrace{\mathbf{e}_1 \otimes \sum_{s=2}^{n} \sum_{J=1}^{J_s^n} (\cdots))}_{(c_1)} \\
&+ 
\underbrace{\mathbf{e}_2 \otimes (\sum_{J=1}^{J_0^n} (\cdots))}_{(a_2)}
+
\underbrace{\mathbf{e}_2 \otimes\sum_{J=1}^{J_1^n} (\cdots))}_{(b_2)}
+ 
\underbrace{\mathbf{e}_2 \otimes \sum_{s=2}^{n} \sum_{J=1}^{J_s^n} (\cdots))}_{(c_2)} \\
&+ 
\underbrace{\mathbf{e}_3 \otimes (\sum_{J=1}^{J_0^n} (\cdots))}_{(a_3)}
+
\underbrace{\mathbf{e}_3 \otimes\sum_{J=1}^{J_1^n} (\cdots))}_{(b_3)}
+ 
\underbrace{\mathbf{e}_3 \otimes \sum_{s=2}^{n} \sum_{J=1}^{J_s^n} (\cdots))}_{(c_3)}.
\end{align*} \\ \\

Suppose 
$(a_1) + (a_2) + (a_3) = (\alpha), \,
(b_1) + (b_2) + (b_3) = (\beta),  \,
(c_1) + (c_2) + (c_3) = (\gamma)$  \\
where ($\alpha$), ($\beta$) and ($\gamma$) will be the resulting sums.

\begin{itemize}
    \item [\textbf{Step 1:}] $(a_1) + (a_2) + (a_3) = (\alpha)$
        \begin{align*}
             &\mathbf{e}_1 \otimes (\sum_{J=1}^{J_0^n} \mathds{H}_J^{(n)} \alpha_{1J}) 
             + 
            \mathbf{e}_2 \otimes (\sum_{J=1}^{J_0^n} \mathds{H}_J^{(n)} \alpha_{2J}) 
            + 
            \mathbf{e}_3 \otimes (\sum_{J=1}^{J_0^n} \mathds{H}_J^{(n)} \alpha_{3J}) \\
            &= 
            \sum_{J=1}^{J_0^n} \mathds{H}_J^{(n)} 
            (\alpha_{1J} \mathbf{e}_1 + \alpha_{2J} \mathbf{e}_2  + \alpha_{3J} \mathbf{e}_3)   \\
             &= 
            \sum_{J=1}^{J_0^n} \mathds{H}_J^{(n)} \textbf{v}_J
    \end{align*}
    
    \item[\textbf{Step 2:}] $(b_1) + (b_2) + (b_3) = (\beta)$ 
        \begin{align*}
            &\mathbf{e}_1 \otimes\sum_{J=1}^{J_1^n} \mathds{H}_J^{(n+1)}[\mathbf{v}_{1J}]  +
            \mathbf{e}_2 \otimes\sum_{J=1}^{J_1^n} \mathds{H}_J^{(n+1)}[\mathbf{v}_{2J}] +
            \mathbf{e}_3 \otimes\sum_{J=1}^{J_1^n} \mathds{H}_J^{(n+1)}[\mathbf{v}_{3J}]\\
            &=
            \sum_{J=1}^{J_1^n} \mathds{H}_J^{(n+1)}(
            \mathbf{e}_1 \otimes \mathbf{v}_{1J} + \mathbf{e}_2 \otimes
            \mathbf{v}_{2J} + \mathbf{e}_3 \otimes
            \mathbf{v}_{3J}) \\
            &=
            \sum_{J=1}^{J_1^n} \mathds{H}_J^{(n+1)}(
            \textbf{e}_k \otimes 
            \mathds{D}_{kJ}^{(1)})\\
            &=
            \sum_{J=1}^{J_1^n} \mathds{H}_J^{(n+1)}(
            \mathds{G}_{J}^{(2)})\\
            &=
            \sum_{J=1}^{J_1^n} \mathds{H}_J^{(n+1)}(\alpha_J \mathbf{1} + \boldsymbol{\epsilon} \cdot \mathbf{v}_J + \mathds{D}_J^{(2)})   
        \end{align*}
        
        Note that the previously established
          component form for $\mathds{G}^{(2)}$ (\ref{finalG}) could not
          be used here, because it is only valid
          for $n > 1$. 
        Therefore the representation for
          a general second order tensor
          was used, as given in the base case.
        
    \item[\textbf{Step 3:}] $(c_1) + (c_2) + (c_3) = (\gamma)$
        \begin{align*}
            &\mathbf{e_1} \otimes \sum_{s=2}^{n} \sum_{J=1}^{J_s^n} \mathds{H}_J^{(n+s)} [\mathds{D}_{1J}^{(s)}] 
            +
            \mathbf{e_2} \otimes \sum_{s=2}^{n} \sum_{J=1}^{J_s^n} \mathds{H}_J^{(n+s)} [\mathds{D}_{2J}^{(s)}] 
            +
            \mathbf{e_3} \otimes \sum_{s=2}^{n} \sum_{J=1}^{J_s^n} \mathds{H}_J^{(n+s)} [\mathds{D}_{3J}^{(s)}]\\
            &=
            \sum_{s=2}^{n} \sum_{J=1}^{J_s^n} \mathds{H}_J^{(n+s)} (
            \mathbf{e}_1 \otimes \mathds{D}_{1J}^{(s)} + \mathbf{e}_2 \otimes \mathds{D}_{2J}^{(s)}+ \mathbf{e}_3 \otimes \mathds{D}_{3J}^{(s)})\\ 
            &=
            \sum_{s=2}^{n} \sum_{J=1}^{J_s^n} \mathds{H}_J^{(n+s)}
            (\textbf{e}_k \otimes \mathds{D}_k^{J^{(s)}})\\
            &=
            \sum_{s=2}^{n} \sum_{J=1}^{J_s^n} \mathds{H}_J^{(n+s)}
            (\mathds{G}_J^{(s+1)})\\ 
            &= 
            \sum_{s=2}^{n} \sum_{J=1}^{J_s^n} \mathds{H}_J^{(n+s)}
            [\mathbf{L}[\mathds{D}^{(s-1)}] + \boldsymbol{\epsilon} \cdot \mathds{D}_J^{(s)} + \mathds{D}^{(s+1)}] 
        \end{align*}
        
        with
        \begin{align}
        L_{k j_1 \cdots j_s l_1\cdots l_{s-1}}D_{l_1 \cdots l_{s-1}} 
        = 
        \frac{2s-1}{s-1}\delta_{\hat{kj_1}}D_{\hat{j}_2 \cdots \hat{j}_s} - \delta_{\hat{j}_1\hat{j}_2}D_{\hat{j}_3 \cdots \hat{j}_sk}
        \end{align}
\end{itemize}

In Step 3 the results of the previous chapter, more precisely the component form for a tensor $\mathds{G}^{(n+1)} = \textbf{e}_k \otimes \mathds{D}_k^{(n+1)}$, was used.

Therefore, the following can be derived:
\begin{align} \label{T(n+1)rekursiv_tensor}
    \mathds{T}^{(n+1)} 
    &=
    (\alpha) + (\beta) + (\gamma) \nonumber  \\  
    &=
    \sum_{J=1}^{J_0^n} \mathds{H}_J^{(n)} \mathbf{v}_J 
    +
    \sum_{J=1}^{J_1^n} \mathds{H}_J^{(n+1)} [\alpha_J \mathbf{1} + \boldsymbol{\epsilon} \cdot \mathbf{v}_J + \mathds{D}_J^{(2)}]
    + 
    \sum_{s=2}^{n} \sum_{J=1}^{J_s^n} \mathds{H}_J^{(n+s)} [\mathbf{L}[ \mathds{D}_J^{(s-1)}] + \boldsymbol{\epsilon} \cdot \mathds{D}_J^{(s)} + \mathds{D}_J^{(s+1)}]
\end{align}

or equivalently written in its component form
\begin{align} \label{T(n+1)rekursiv_component}
    T_{k i_1 \cdots i_n} 
    &= \sum_{J=1}^{J_0^n} H_{i_1 \cdots i_n}^{J} v_k^J
    + \sum_{J=1}^{J_1^n} H_{i_1 \cdots i_n j}^{J} [\alpha^J \delta_{jk} + \epsilon_{jkt}v_t^J + D_{jk}^J]\\ 
    &+ \sum_{s=2}^n \sum_{J=1}^{J_s^n} H_{i_1 \cdots i_n j_1 \cdots j_s}^{J} [L_{k j_1 \cdots j_s l_1 \cdots l_{s-1}} D_{l_1 \cdots l_{s-1}} + \epsilon_{k t \hat{j}_1}D_{\hat{j}_2 \cdots \hat{j}_s t}^J + D_{k j_1 \cdots j_s}] 
\end{align}

 with
\begin{align*}
    L_{k j_1 \cdots j_s l_1\cdots l_{s-1}}D_{l_1 \cdots l_{s-1}} 
    = 
    \frac{2s-1}{s-1}\delta_{\hat{kj_1}}D_{\hat{j}_2 \cdots \hat{j}_s} - \delta_{\hat{j}_1\hat{j}_2}D_{\hat{j}_3 \cdots \hat{j}_sk}.
    \end{align*}

This recursive representation of $\mathds{T}^{(n+1)}$ was established by Zou et al. \cite{zou2001orthogonal}. \\

\textbf{Representation 2}

In contrast, another representation of $\mathds{T}^{(n+1)}$ can be derived by using the deviatoric decomposition and writing $(n+1)$ instead of $n$ by
\begin{align} \label{T(n+1)_normal}
    \mathds{T}^{(n+1)} 
    &= 
    \sum_{J=1}^{J_0^{n+1}} \alpha_{J}\mathds{H}_J^{(n+1)} 
    +
    \sum_{J=1}^{J_1^{n+1}} \mathds{H}_J^{((n+1)+1)} [\mathbf{v}_{J}] 
    + 
    \sum_{s=2}^{n+1} \sum_{J=1}^{J_s^{n+1}} \mathds{H}_J^{((n+1)+s)} [\mathds{D}_{J}^{(s)}].\\\nonumber
\end{align}
\textbf{Calculation of the Number of Independent $s$th Order Deviatoric Tensors $J_s^n$}

By comparing these two representations (\ref{T(n+1)rekursiv_tensor}) and (\ref{T(n+1)_normal}) it becomes obvious, that both are a sum of the same irreducible parts containing $s$th deviatoric tensors, because the hemitropic tensor $\mathds{H}$ is only a linear combination of tensor products of $\mathds{1}$ and $\boldsymbol{\varepsilon}$. Therefore, the representations are identical, if the number of $s$th order deviatoric tensors $J_s^n$ is the same in both representations.

By counting the deviatoric tensors of the same order in representation 1, the following number property for the number $J_s^{n+1}$ is gained, as done by Zou et al. \cite{zou2001orthogonal}.
\begin{align*}
    J_s^{(n+1)} 
    = \begin{cases}
    J_{s-1}^n + J_s^n + J_{s+1}^n & , 1 \le s \le n 
    \\
    J_1^n & , s = 0 
    \\
    J_n^n & , s = n+1
    \end{cases}
\end{align*}

According to Zou et al. \cite{zou2001orthogonal} this property and the recursive form of 
  representation 1 can be used repeatedly
  to find the calculation for the number
  $J_s^n, 0 \le s \le n$:
  \begin{align}\label{J(s,n)}
     J_s^n = \begin{bmatrix} n \\ s \end{bmatrix}  
     - 
     \begin{bmatrix} n \\ s+1 \end{bmatrix}
  \end{align}
where $\begin{bmatrix} n \\ s \end{bmatrix}, 0 \le s \le n$, represents the trinomial coefficients of $x^s$ in the expansion of $(1 + x + \frac{1}{x})^n$. 
This statement can be proven by induction. 
For this properties of the trinomial coefficient, such as its recurisve calculation,
\begin{align}\label{recursive_trinomial}
        \begin{bmatrix} n+1 \\ s \end{bmatrix}   
        =
        \begin{bmatrix} n \\ s-1 \end{bmatrix}
        + 
        \begin{bmatrix} n \\ s \end{bmatrix}
        +
        \begin{bmatrix} n \\ s+1 \end{bmatrix}, n \ge 0
\end{align}
will be used.
Noting $J_0^0 = 1$ the base case is trivial.
Assuming the statement (\ref{J(s,n)}) is true for a certain integer $n$, the statement will now be proven for $n+1$.
Since there are three cases for the calculation of $J_s^{n+1}$, the inductive step will also include three cases.

\begin{itemize}
    \item \textbf{s = 0}
        \begin{align*}
            \begin{bmatrix} n+1 \\ 0 \end{bmatrix}
            -
            \begin{bmatrix} n+1 \\ 1 \end{bmatrix}
            =
            J_0^{n+1} 
            = 
            J_1^n
            = 
            \begin{bmatrix} n \\ 1 \end{bmatrix}
            - 
            \begin{bmatrix} n \\ 2 \end{bmatrix}
        \end{align*}
        
    \item \textbf{s = n+1}
        \begin{align*}
            \begin{bmatrix} n+1 \\ n + 1\end{bmatrix}
            - 
            \begin{bmatrix} n+1 \\ n + 2\end{bmatrix}
            =
            J_{n+1}^{n+1} 
            =
            J_n^n
            = 
            \begin{bmatrix} n \\ n \end{bmatrix}
            - 
            \begin{bmatrix} n \\ n+1 \end{bmatrix}
        \end{align*}
        
    \item $\mathbf{1 \le s \le n}$
        \begin{align*}
            \begin{bmatrix} n+1 \\ s \end{bmatrix}
            -
            \begin{bmatrix} n+1 \\ s+1 \end{bmatrix}
            =
            J_s^{n+1}
            =
            J_{s-1}^n + J_s^n + J_{s-1}^n
            =
            \begin{bmatrix} n \\ s-1 \end{bmatrix}
            -
            \begin{bmatrix} n \\ s+2 \end{bmatrix}
        \end{align*}
\end{itemize}

Applying the recursive calculation for the    trinomial coefficient  
  (\ref{recursive_trinomial}), the statement is proven. 

All in all it can be stated, that representation 1 and 2 are identical if and only if the number $J_s^n, 0 \le s \le n,$ is calculated as in
(\ref{J(s,n)}). 
An example for the number $J_s^n$ up to $n=6$ is given in table 1.
\begin{table} [h]
\centering
\begin{tabular}{|M{0.5cm}||M{0.5cm}|M{0.5cm}|M{0.5cm}|M{0.5cm}|M{0.5cm}| M{0.5cm}|M{0.5cm}|} 
\hline 
\vspace{0.1cm}n & \vspace{0.1cm}$J_0^n$ \vspace{0.1cm}&\vspace{0.1cm} $J_1^n$
\vspace{0.1cm}&\vspace{0.1cm} $J_2^n$ \vspace{0.1cm}& \vspace{0.1cm}$J_3^n$ \vspace{0.1cm}&\vspace{0.1cm} $J_4^n$ \vspace{0.1cm}&\vspace{0.1cm} $J_5^n$ \vspace{0.1cm}& \vspace{0.1cm}$J_6^n$\\
\hline \hline 
\vspace{0.1cm}0 & \vspace{0.1cm}1 &  &  &  &  &  & \\
\hline
\vspace{0.1cm}1 & \vspace{0.1cm}0 &\vspace{0.1cm}1  &  &  &  &  & \\
\hline
\vspace{0.1cm}2 & \vspace{0.1cm}1 & \vspace{0.1cm}1 &\vspace{0.1cm}1  &  &  &  & \\
\hline
\vspace{0.1cm}3 &\vspace{0.1cm}1  &\vspace{0.1cm} 3 &\vspace{0.1cm} 2 &\vspace{0.1cm} 1 &  & & \\
\hline
\vspace{0.1cm}4 & \vspace{0.1cm}3 & \vspace{0.1cm}6 &\vspace{0.1cm}6  & \vspace{0.1cm}3 &\vspace{0.1cm}1  &  & \\
\hline
\vspace{0.1cm}5 & \vspace{0.1cm}6 & \vspace{0.1cm}15 & \vspace{0.1cm}15 &\vspace{0.1cm}10  & \vspace{0.1cm}4 & \vspace{0.1cm}1 & \\
\hline
\vspace{0.1cm}6 & \vspace{0.1cm}15 & \vspace{0.1cm}36 &\vspace{0.1cm}40  & \vspace{0.1cm}29 & \vspace{0.1cm}15 &\vspace{0.1cm} 5 &\vspace{0.1cm}1 \\
\hline
\end{tabular} 
\caption{\vspace{0.5cm}The numbers of independent deviatoric tensors in the irreducible decompositions of generic tensors in 3D.}
\end{table}

Next, it needs to be proven that the deviatoric tensors in 
  (\ref{ZerlegungT(n)}) are independent and therefore are no linear combination of one another. 
Following the idea by Zou et al. \cite{zou2001orthogonal}, this is true
  if
\begin{align*}
    3^n 
    =
    \sum_{s = 0}^n (1+2s)J_s^n.
\end{align*}
This statement will also be proven by induction. 
Since $J_0^0 = 1 = 3^0$ the base case is trivial. 
For the inductive step follows:
\begin{align*}
    \sum_{s = 0}^{n+1} (1+2s)J_s^{n+1}
    &=
    (2n+3)J_{n+1}^{n+1} + J_0^{n+1} +
    \sum_{s = 1}^{n}(1+2s)J_s^{n+1}\\
    &=
    (2n+3)J_n^n + J_1^n +
    \sum_{s = 1}^n (1+2s)(J_{s-1}^n + J_s^n + J_{s+1}^n) \\
    &= (2n+3)J_n^n + J_1^n + 
    \sum_{s = 1}^{n} (1+2s)J_{s-1}^n +
    \sum_{s = 1}^{n} (1+2s)J_s^n + 
    \sum_{s = 1}^{n} (1+2s)J_{s+1}^n \\
    &= (2n+3)J_n^n + J_1^n 
    + \sum_{s = 0}^{n-1} (1+2(s+1))J_s^n 
    + \sum_{s = 1}^{n} (1+2s)J_s^n 
    + \sum_{s = 2}^{n+1} (1+2(s-1))J_s^n \\
    &= (2n+3)J_n^n + J_1^n
    + (\sum_{s = 0}^{n} (1+2(s+1))J_s^n) - (2n+3)J_n^n \\
    & \quad
    + (\sum_{s = 0}^{n} (1+2s)J_s^n) -J_0^n 
    + (\sum_{s = 0}^{n} (1+2(s-1)) J_s^n) - J_1^n + J_0^n\\
    &= 3\sum_{s = 0}^{n} (1+2s)J_s^n\\
    &= 3\cdot 3^n \\
    &= 3^{n+1}
    \end{align*}
Thus, all deviatoric tensors in (\ref{ZerlegungT(n)}) are independent.

It remains to prove that the decomposition
  (\ref{ZerlegungT(n)}) is orthogonal.
A decomposition is orthogonal, if the sets of
  $\mathds{H}_J^{n+s}[\mathds{D}^{(s)}]$ for all $\mathds{D}^{(s)}\in \mathcal{D}^{(s)}$ 
  are mutually orthogonal for different pairs $(s,J)$.  
In their paper, Zou et al. \cite{zou2001orthogonal} explained the prove by induction of this statement
  in detail for 2 dimensions.
As described by them, the orthogonality can be derived similarly in 3 dimensions.

\section{Examples}
\subsection{General Tensors}
In the following the recursive formula 
\begin{align*} 
    T_{k i_1 \cdots i_n} 
    &= \sum_{J=1}^{J_0^n} H_{i_1 \cdots i_n}^{J} v_k^J
    + \sum_{J=1}^{J_1^n} H_{i_1 \cdots i_n j}^{J} [\alpha^J \delta_{jk} + \epsilon_{jkt}v_t^J + D_{jk}^J]\\ 
    &+ \sum_{s=2}^n \sum_{J=1}^{J_s^n} H_{i_1 \cdots i_n j_1 \cdots j_s}^{J} [L_{k j_1 \cdots j_s l_1 \cdots l_{s-1}} D_{l_1 \cdots l_{s-1}}^J + \epsilon_{k t \hat{j}_1}D_{\hat{j}_2 \cdots \hat{j}_s t}^J + D_{k j_1 \cdots j_s}^J] 
\end{align*}

 with
\begin{align}\label{LDcomponent}
    L_{k j_1 \cdots j_s l_1\cdots l_{s-1}}D_{l_1 \cdots l_{s-1}} 
    = 
    \frac{2s-1}{s-1}\delta_{\hat{kj_1}}D_{\hat{j}_2 \cdots \hat{j}_s} - \delta_{\hat{j}_1\hat{j}_2}D_{\hat{j}_3 \cdots \hat{j}_sk}
    \end{align}
  will be used to find the deviatoric decomposition 
  for arbitrary tensors of order three and four 
  building on the well-known decomposition of a second order tensor (\ref{decomp2order}).
  
  It is important to explain certain notations used in the following.
\begin{itemize}
    \item  If the component form of a $n$th-order tensor is $T_{i_1 \cdots i_n}$, 
    then, the recursive formula indicates, that the $(n+1)$th-order tensor gains a
    new index variable $k$, which is put in front of the index.
  Therefore, the component form of a $(n+1)$th order tensor would be $T_{k i_1 \cdots i_n}$.
  In the following, the "new index", which is added through the recursive formula, will always be $i$.
  This allows for the component form of the examples to 
  be $T_{i j}$, $T_{i j k}$ and $T_{i j k l}$, 
  rather than
  $T_{k i }$, $T_{k i j}$ and $T_{k i j l}$.
  
  \item Due to the definition of the recursive formula, the sum $\sum_{J = 1}^{J_s^n}, 1 \le s \le n,$
  suggests, that there are several deviatoric tensors with the same index $J$.
  For example: Since every sum starts at $J = 1$, there would be 3 first order
  deviatoric tensors $v_k^1$ for $n = 4$.
  As proven in the previous section, all deviatoric tensors are independent and therefore not equal.
  This allows in the following to adjust the count of the index $J$ appropriately.
  So instead of $v_k^1$,  $v_k^{1'}$ and  $v_k^{1''}$, they will be denoted as
   $v_k^1$,  $v_k^2$ and  $v_k^3$.
   
  \item The calculation of the final form will be done in the follwing way. 
  At first the recursive formula will be written down for the given $n$.
  Next, the hemitropic tensors will be inserted as derived in the $(n-1)$th case.
  After that, the summands will be sorted by the order of their deviatoric tensor.
  At last the contraction variables will be adjusted, so that not too many different index variables will be used in the final form.
  
  \item The final forms given here will differ from the results given by Zou et al. \cite{zou2001orthogonal} by 
  certain index permutations within the irreducible parts.
  Noting that this will only result in a different linear combination, both results are correct.
  
  \item According to Zou et al. \cite{zou2001orthogonal} the tensor $L_{i j \cdots k}$ is defined by
  the tensor contraction given in line (\ref{LDcomponent}).
  Therefore, for $L_{i j k l}$ yields
  \begin{align*}
      L_{i j k l} = \frac{3}{2}(\delta_{i j}\delta_{k l} + \delta_{i k}\delta_{j l}) - \delta_{i l }\delta_{j k}
  \end{align*}
  
\end{itemize}

\begin{enumerate}
    \item[\boldsymbol{$n = 2$}] 
    The deviatoric decomposition is, as explained in (\ref{decomp2order}):
    \begin{align*}
        T_{ij} = \alpha\delta_{ij} + \epsilon_{ijs}v_s + D_{ij}.
    \end{align*}
    For the number $J_s^n, 0 \le s \le n,$ of the independent $s$th-order
    deviatoric tensors yields: 
    \begin{align*}
    J_0^2 = 1,\quad  J_1^2 = 1,\quad J_2^2 = 1.
    \end{align*}
    Therefore, the hemitropic tensors are described as:
    \begin{itemize}
        \item $H_{j k}^{1} = \delta_{j k }$
        \item $H_{j k s}^{1} = \epsilon_{j k s}$
        \item $H_{j k s t}^1 = \delta_{j s}\delta_{k t}$
    \end{itemize}
    \item[\boldsymbol{$n = 3$}] 
    Using the recursive formula:
    \begin{align*} 
    T_{i j k} 
    &= \sum_{J=1}^{J_0^{n-1}= J_0^2 = 1} 
            H_{j k}^{J} v_i^J \\
        &\quad + \sum_{J=1}^{J_1^{n-1} = J_1^2 = 1} 
            H_{j k s}^{J} 
            [\alpha^J \delta_{s i} 
            + \epsilon_{s i t}v_t^J 
            + D_{s i}^J]\\ 
        &\quad + \sum_{J=1}^{J_2^{n-1} = J_2^2 = 1}  
            H_{i j s t} 
            [L_{i s t u } D_{u}^J 
            + \epsilon_{i v \hat{s}}D_{\hat{t} v}^J 
            + D_{i s t}^J] \\
    &= \delta_{j k}v_i^1 
        + \alpha \delta_{s i}\epsilon_{j k s} 
        + \epsilon_{j k s} \epsilon_{s i t} v_t^2 
        + \epsilon_{j k s}D_{s i}^1
        + \delta_{j s}\delta_{k t}L_{i s t u}v_u^3 
        + \delta_{j s}\delta_{k t}\epsilon_{i v \hat{s}}D_{\hat{t}v}^2 
        + \delta_{j s}\delta_{k t}D_{i s t}  \\
    &= \epsilon_{j k i} \alpha
      + \{ 
        \delta_{j k }v_i^1 
        + \epsilon_{j k s}\epsilon_{s i t} v_t^2 
        + L_{i j k u}v_u^3  
        \}
      + \{ 
        \epsilon_{j k s}D_{s i}^1 
        + \epsilon_{iv\hat{j}}D_{\hat{k}v}^2   
        \}
      + D_{i j k}\\
    &= \epsilon_{j k i} \alpha
      + \{ 
        \delta_{j k }v_i^1 
        + \epsilon_{j k t}\epsilon_{t i s} v_s^2 
        + L_{i j k s}v_s^3   
        \}
      + \{ 
        \epsilon_{j k s}D_{s i}^1 
        + \epsilon_{is\hat{j}}D_{\hat{k}s}^2   
        \}
      + D_{i j k}
    \end{align*}
    For the number $J_s^n, 0 \le s \le n,$ of the independent $s$th-order
    deviatoric tensors yields: 
    \begin{align*}
        J_0^3 = 1,\quad J_1^3 = 3,\quad J_2^3 = 2,\quad J_3^3 = 1
    \end{align*}
    Therefore, the hemitropic tensors are described as 
    \begin{itemize}
        \item $H_{j k l}^1 = \epsilon_{k l j}$
        \item $H_{j k l s}^1 = \delta_{s j}\delta_{k l }$
        \item $H_{j k l s}^2 =  \epsilon_{k l t}\epsilon_{t j s}$
        \item $H_{j k l s}^3 =  L_{j k l s}$
        \item $H_{j k l s t}^1 = \delta_{j t}\epsilon_{k l s}$ 
        \item $H_{j k l s t}^2 =  \delta_{\hat{l} t} \epsilon_{j s \hat{k}}$ 
        \item $H_{j k l s t u}^1 = \delta_{j s}\delta_{k t}\delta_{l u}$
    \end{itemize}
    
    \item[\boldsymbol{$n = 4$}]
    Using the recursive formula:
    \begin{align*} 
    T_{i j k l} 
    &= \sum_{J=1}^{J_0^{n-1} = J_0^3 = 1} 
            H_{j k l}^{J} v_i^J\\
        &\quad + \sum_{J=1}^{J_1^{n-1} = J_1^3 = 3} 
            H_{j k l s}^{J} 
            [\alpha^J \delta_{s i} 
            + \epsilon_{s i t}v_t^J 
            + D_{s i}^J]\\ 
        &\quad + \sum_{J=1}^{J_2^{n-1} = J_2^3 = 2} 
            H_{j k l s t}^{J} 
            [L_{i s t x_1} v_{x_1 }^J
            + \epsilon_{i v \hat{s}}D_{\hat{t} v}^J 
            + D_{i s t}^J] \\
        &\quad + \sum_{J=1}^{J_3^{n-1}=J_3^3=1} 
            H_{j k l s t u}^{J}
            [L_{i s t u x_1 x_2} D_{x_1 x_2} ^J
            + \epsilon_{i v \hat{s}}D_{\hat{t} \hat{u} v}^J 
            + D_{i s t u}^J] \\
    &= \epsilon_{k l j} v_i^1 \\ 
        &\quad  +   \delta_{s j}\delta_{k l } \alpha^1 \delta_{s i} 
                + \delta_{s j}\delta_{k l } \epsilon_{s i t}v_t^2 
                + \delta_{s j}\delta_{k l } D_{s i}^1 \\ 
        &\quad  +   \epsilon_{k l t}\epsilon_{t j s} \alpha^2 \delta_{s i} 
                + \epsilon_{k l t}\epsilon_{t j s} \epsilon_{s i t}v_t^3  
                + \epsilon_{k l t}\epsilon_{t j s} D_{s i}^2 \\ 
        &\quad  + L_{j k l s} \alpha^3 \delta_{s i} 
                + L_{j k l s} \epsilon_{s i t}v_t^4 
                + L_{j k l s} D_{s i}^3 \\ 
        &\quad  + \delta_{j t}\epsilon_{k l s} L_{i s t x_1} v_{x_1 }^5 
                + \delta_{j t}\epsilon_{k l s} \epsilon_{i v \hat{s}}D_{\hat{t} v}^4
                + \delta_{j t}\epsilon_{k l s} D_{i s t}^1 \\
        &\quad  + \delta_{\hat{l} t} \epsilon_{j s \hat{k}} L_{i s t x_1} v_{x_1 }^6 
                + \delta_{\hat{l} t} \epsilon_{j s \hat{k}} \epsilon_{i v \hat{s}}D_{\hat{t} v}^5 
                + \delta_{\hat{l} t} \epsilon_{j s \hat{k}} D_{i s t}^2  \\ 
        &\quad  + \delta_{j s}\delta_{k t}\delta_{l u} L_{i s t u x_1 x_2} D_{x_1 x_2} ^6 
                + \delta_{j s}\delta_{k t}\delta_{l u} \epsilon_{i v \hat{s}}D_{\hat{t} \hat{u} v}^3 
                + \delta_{j s}\delta_{k t}\delta_{l u} D_{i s t u}^1
    \end{align*}
    
    \begin{align*}
    &= \{
                  \delta_{i j}\delta_{k l } \alpha^1 
                + \epsilon_{k l t}\epsilon_{t j i} \alpha^2  
                +  L_{j k l i} \alpha^3 
        \} \\
        &\quad  + 
        \{  
                   \epsilon_{k l j} v_i^1   
                +  \delta_{k l } \epsilon_{j i t}v_t^2 
                + \epsilon_{k l t}\epsilon_{t j s} \epsilon_{s i t}v_t^3
                + L_{j k l s} \epsilon_{s i t}v_t^4 
                + \epsilon_{k l s} L_{i s j x_1} v_{x_1 }^5
                + \epsilon_{j s \hat{k}} L_{i s \hat{l} x_1} v_{x_1 }^6
        \} \\
        &\quad +
        \{
                 \delta_{k l } D_{j i}^1
                + \epsilon_{k l t}\epsilon_{t j s} D_{s i}^2  
                + L_{j k l s} D_{s i}^3
                + \epsilon_{k l s} \epsilon_{i v \hat{s}}D_{\hat{j} v}^4
                + \epsilon_{j s \hat{k}} \epsilon_{i v \hat{s}}D_{\hat{l} v}^5
                + L_{i j k l x_1 x_2} D_{x_1 x_2} ^6
        \}\\
        &\quad +
        \{
                \epsilon_{k l s} D_{i s j}^1
                + \epsilon_{j s \hat{k}} D_{i s \hat{l}}^2 
                + \epsilon_{i v \hat{j}}D_{\hat{k} \hat{l} v}^3 
        \}\\
        &\quad +  D_{i j k l}^1\\
    &= \{
                  \delta_{i j}\delta_{k l } \alpha^1 
                + \epsilon_{k l t}\epsilon_{t j i} \alpha^2  
                +  L_{j k l i} \alpha^3 
        \} \\
        &\quad  + 
        \{  
                   \epsilon_{k l j} v_i^1   
                +  \delta_{k l } \epsilon_{j i s}v_s^2 
                + \epsilon_{k l t}\epsilon_{t j s} \epsilon_{s i t}v_t^3
                + L_{j k l s} \epsilon_{s i t}v_t^4 
                + \epsilon_{k l s} L_{i s j t} v_{t }^5
                + \epsilon_{j s \hat{k}} L_{i s \hat{l} t} v_{t }^6
        \} \\
        &\quad +
        \{
                 \delta_{k l } D_{j i}^1
                + \epsilon_{k l t}\epsilon_{t j s} D_{s i}^2  
                + L_{j k l s} D_{s i}^3
                + \epsilon_{k l s} \epsilon_{i t \hat{s}}D_{\hat{j} t}^4
                + \epsilon_{j s \hat{k}} \epsilon_{i t \hat{s}}D_{\hat{l} t}^5
                + L_{i j k l s t} D_{s t} ^6
        \}\\
        &\quad +
        \{
                \epsilon_{k l s} D_{i s j}^1
                + \epsilon_{j s \hat{k}} D_{i s \hat{l}}^2 
                + \epsilon_{i s \hat{j}}D_{\hat{k} \hat{l} s}^3 
        \}\\
        &\quad +  D_{i j k l}^1
     \end{align*}
     For the number $J_s^n, 0 \le s \le n,$ of the independent $s$th-order
    deviatoric tensors yields: 
    \begin{align*}
       J_0^4 = 3,\quad J_1^4 = 6,\quad J_2^4 = 6,\quad J_4^4 = 3,\quad J_4^4 = 1
    \end{align*}
\end{enumerate}
Following the same procedure the deviatoric decomposition for tensors of any order $n$ can be derived.
Zou et al. \cite{zou2001orthogonal} presented the deviatoric decomposition up to order 5.

\subsection{Third Order Coupling Tensor}
The third order coupling tensor $\mathds{H}^{(3)}$ is defined as a three dimensional
  third order tensor, which represents the sensitivity of the Piola-Kirchhoff-type stress tensor $\mathds{S}^{(2)}$ with respect to the electric field vector $\mathds{E}^{(1)}$.
It displays the following symmetry
\begin{align*}
    H_{i j k} = H_{j i k}.
\end{align*}
Note, that in this chapter $\mathds{H}^{(3)}$, or $H_{i j k}$, denotes the coupling tensor and not a hemitropic tensor as in the previous sections.
Hergl \cite{herglanalysis} gave a specific calculation method to determine 
  the component forms for the deviatoric tensors.
She used the recursive formula to determine the deviatoric decomposition 
  for a general third order tensor, as demonstrated in this work.
After stating the component forms for $T_{iii}, T_{ijj}, T_{iij}, T_{iji}$ and $T_{i j k}$
  she generated an equation system to calculate equations for the deviatoric tensors
  by calculating the double contraction with the second order identity tensor $\delta_{i j}$
  or the permutation tensor $\epsilon_{i j k}$.
Applying the mentioned symmetry of the coupling tensor, she gained the following result.

For the zeroth order deviator $\alpha$, the first order deviator $\textbf{v}^1$,
  and the second order deviator $D_{i j}^2$ it is
\begin{align*}
\alpha = 0, \quad \textbf{v}^1 = \frac{5}{2} \textbf{v}^3 - \textbf{v}^2, \quad D_{ij}^2 = -\frac{2}{3}D_{ij}^1.
\end{align*}

Thus, the zeroth order deviator vanishes, and one first order deviator and 
  one second order deviator are dependent of the other deviators.
This means that the coupling tensor can be represented by
  four deviators: two of order one, one of order two
  and one of order three.
For the associated coefficients for the deviatoric tensors follows:\\


For the first order deviators $\textbf{v}^2$ and $\textbf{v}^3$, the second order deviator
  $D_{i j}^1$, and the third order deviator $D_{i j k}$ it is
\begin{equation}
\begin{split}
	&\textbf{v}^2 = \frac{1}{4} \begin{pmatrix}
	H_{221} - H_{122} + H_{331} - H_{133}\\
	H_{112} - H_{121} + H_{332} - H_{233} \\
	H_{133} - H_{131} + H_{223} - H_{232}
	\end{pmatrix}\\
	&\textbf{v}^3 = \frac{1}{30} \begin{pmatrix}
	4H_{111} + H_{122} + H_{133} + 3H_{221} +3H_{331}\\
	4H_{222} + H_{121} + H_{233} + 3H_{112} +3H_{332}\\
	4H_{333} + H_{131} + H_{232} + 3H_{113} +3H_{223}
	\end{pmatrix},
\end{split}
\end{equation}
and
\begin{equation}
\begin{split}
&D_{11}^1 = \frac{1}{2}\left( H_{123} - H_{132}\right)\\
&D_{22}^1 = \frac{1}{2}\left( H_{231} - H_{123}\right)\\
&D_{33}^1 = \frac{1}{2}\left( H_{132} - H_{231}\right)\\
&D_{12}^1 = \frac{1}{4}\left(-H_{113}+H_{131}+H_{223}-H_{232}\right)\\
&D_{13}^1 = \frac{1}{4}\left( H_{112}-H_{121}+H_{233}-H_{332}\right)\\
&D_{23}^1 = \frac{1}{4}\left( H_{122}-H_{221}-H_{133}+H_{331}\right)
\end{split}
\end{equation}
and 
\begin{equation}
\begin{split}
&D_{111} = \frac{2}{5} H_{111} - \frac{2}{5}H_{122} - \frac{2}{5}H_{133} - \frac{1}{5}H_{221} -\frac{1}{5}H_{331}\\
&D_{222} = \frac{2}{5}H_{222}-\frac{2}{5}H_{211}-\frac{2}{5}H_{233}-\frac{1}{5}H_{112} - \frac{1}{5} H_{332}\\
&D_{333} = \frac{2}{5}H_{333}-\frac{2}{5}H_{322}-\frac{2}{5}H_{311}-\frac{1}{5}H_{113}-\frac{1}{5}H_{223}\\
&D_{122} = \frac{8}{15}H_{122}-\frac{1}{5}H_{111}-\frac{2}{15}H_{133}+\frac{4}{15}H_{221}-\frac{1}{15}H_{331}\\
&D_{133} = \frac{8}{15}H_{133}-\frac{1}{5}H_{111}-\frac{1}{15}H_{221}+\frac{4}{15}H_{331}-\frac{2}{15}H_{212}\\
&D_{211} = \frac{8}{15}H_{211}-\frac{1}{5}H_{222}-\frac{2}{15}H_{233}+ \frac{4}{15}H_{112}-\frac{1}{15}H_{332}\\
&D_{233} = \frac{8}{15}H_{233}-\frac{2}{15}H_{211}-\frac{1}{5}H_{222} -\frac{1}{15}H_{112}+ \frac{4}{15}H_{332}\\
&D_{311} = \frac{8}{15}H_{311}-\frac{2}{15}H_{322}-\frac{1}{5}H_{333}+\frac{4}{15}H_{113}-\frac{1}{15}H_{223}\\
&D_{322} = \frac{8}{15}H_{322}-\frac{2}{15}H_{311}-\frac{1}{5}H_{333}-\frac{1}{15}H_{113} + \frac{4}{15}H_{223}\\
&D_{123}= \frac{1}{3}H_{123} + \frac{1}{3}H_{132} + \frac{1}{3}H_{231}.
\end{split}
\end{equation}
Thus, the orthogonal irreducible decomposition of the third order coupling tensor can be represented as 
\begin{align*}
    H_{i j k}
    &= 
      \{ 
        \epsilon_{j k t}\epsilon_{t i s} v_s^2
         - \delta_{j k}v_i^2
        + \frac{5}{2}\delta_{j k} v_i^3
        + [\frac{3}{2}(\delta_{i j}\delta_{k s} + \delta_{i k}\delta_{j s}) - \delta_{i s}\delta_{j k}] v_s^3
        \}\\
      &+ \{ 
        \epsilon_{j k s}D_{s i}^1 
        - \frac{1}{3} (\epsilon_{i s j}D_{k s}^1 
        + \epsilon_{i s k}D_{j s}^1)
        \}
      + D_{i j k}
\end{align*}

\subsection{Fourth Order Stiffness Tensor}
The stiffness tensor $\mathds{C}^{(4)}$ is defined as a three dimensional fourth order tensor, 
  which describes
  the relation between changes of stress and strain.
It displays two kinds of symmetries, the so-called minor symmetries
    \begin{align*}
     C_{i j k l} = C_{j i k l } = C_{i j l k}    
    \end{align*}
    and the so-called major symmetry
    \begin{align*}
     C_{i j k l} = C_{k l i j}     
    \end{align*}
Due to this fact, deviatoric tensors, which are contracted with a hemitropic tensor involving 
  the permutation tensor $\boldsymbol{\varepsilon}$ vanish in the decomposition.
Therefore, the stiffness tensor can be represented by
  two zeroth order deviators, two second order deviators and one fourth order deviator.
The following irreducible devaitoric decomposition for the stiffness tensor can be
  found in various literature sources \cite{zou2001orthogonal,
herglanalysis,
hergl2020introduction,
zou2013identification}.
In this form the zeroth order deviators are the so-called Lamé-coefficients, which are well
  known in the engineering community.
However, written in this form, the decomposition is not a direct result from the recursive formula,
  but a linear combination of it.
In result, it looses its orthogonality between the two irreducible parts containing the scalars
  and between the two irreducible parts containing the second order deviatoric tensors.
\begin{align}
    C_{i j k l}
    &=
    \lambda \delta_{i j} \delta_{k l} + \mu(\delta_{i k}\delta_{j l} + \delta_{i l}\delta_{j k}) \\
    &+ \{ \delta_{i j }D_{k l}^1 + \delta_{k l}D_{i j}^1 \} 
     + \{ \delta_{i k }D_{j l}^2 + \delta_{i l }D_{j k}^2 + \delta_{j k }D_{i l}^2 + \delta_{j l }D_{i k}^2\} \nonumber \\
    &+ D_{i j k l} \nonumber
\end{align}
Zou et al. \cite{zou2013identification} stated an explicit calculation for the coefficients of the involved 
  deviatoric tensors
\begin{align*}
    \lambda &= \frac{1}{15}(2C_{i i k k} - C_{i k i k}) \\
    \mu &= \frac{1}{30}(3C_{i k i k} - C_{i i k k})\\
    D_{i j}^1 &= \frac{5}{7}(C_{k k i j} - \frac{1}{3}C_{k k l l}\delta_{i j})
      - \frac{4}{7}(C_{k i k j} - \frac{1}{3}C_{k l k l}\delta_{i j}) \\
    D_{i j}^2 &= \frac{3}{7}(C_{k i k j} - \frac{1}{3} C_{k l k l} \delta_{i j})
      - \frac{2}{7}(C_{k k i j} - \frac{1}{3}C_{k k l l} \delta_{i j})\\
    D_{i j k l} &= C_{i j k l } 
    - (
    \lambda \delta_{i j} \delta_{k l} + \mu(\delta_{i k}\delta_{j l} + \delta_{i l}\delta_{j k}) 
    + \{ \delta_{i j }D_{k l}^1 + \delta_{k l}D_{i j}^1 \} 
     + \{ \delta_{i k }D_{j l}^2 + \delta_{i l }D_{j k}^2 + \delta_{j k }D_{i l}^2 + \delta_{j l }D_{i k}^2\}
    )
\end{align*}

\section{Conclusion}
The main goal of this work was to gather numerous information of
  the deviatoric decomposition of an arbitrary
  $n$th-order tensor in three dimensions and to confirm
  details of the found statements.
  
To prove the existence of the deviatoric decomposition mathematical induction was used.
It was seen, that the key was to
  examine the deviatoric decomposition of tensors $\mathds{G}^{(n+1)} \in \mathcal{V}\otimes \mathcal{D}^{(n)}$.
A component form for $\mathds{G}^{(n+1)}$ was found by analyzing certain linear combinations.
With this knowledge two representations for $\mathds{T}^{(n+1)}$ could be stated.
By comparing them a calculation for the number $J_s^n, 0 \le s \le n,$ was found
  and the prove by induction was completed.
In the last chapter the recursive formula was used to
  formulate the deviatoric decomposition 
  for arbitrary tensors up to order 4.
These formulas were then used to describe the decomposition of
  the third order coupling tensor and the fourth order stiffness tensor.
Additionally the a calculation for the coefficients of the involved deviatoric tensors were given.

Especially the recursive representation of the deviatoric decomposition is a powerful
  tool, whose importance needs to be highlighted and meaning investigated even further.

The fact, that deviatoric tensors can further be decomposed into multipoles
  has a great impact, for example, on finding and 
  analyzing symmetry types of certain materials.
According to  
  Zou et al. \cite{zou2013identification} 
  the anisotropy type of the stiffness tensor can be determined by
  analyzing the intersection of the symmetry planes of each deviatoric tensor.
These symmetry planes can be calculated using the multipole decomposition.

Evidently, there are many more details about this decomposition, that have
  not been discussed in this work nor have even been discovered yet.
It would be desirable to motivate other researchers in this field to 
  explore the meaning of this decomposition and the multipoles,
  since it represents a great method to 
  decompose an arbitrary $n$th-order tensor,
  without any particular symmetry type in three dimensions.

\bibliographystyle{unsrt}  

\bibliography{main}

\end{document}